\documentclass[10pt, a4paper, twocolumn]{article} 

\usepackage{xcolor}
\usepackage{array}
\usepackage{amssymb}
\usepackage{hyperref}
\usepackage{colortbl}
\usepackage[
  margin=1.5cm,
  includefoot,
  footskip=30pt,
]{geometry}

\usepackage{amsmath}
\newcounter{rownumber}
\newcounter{rownumbertwo}
\newcommand\rownb{\stepcounter{rownumber}\arabic{rownumber}}

\let\OLDthebibliography\thebibliography
\renewcommand\thebibliography[1]{
  \OLDthebibliography{#1}
  \setlength{\parskip}{0pt}
  \setlength{\itemsep}{0pt plus 1ex}
}

\usepackage[resetlabels,labeled]{multibib}
\newcites{Supp}{Supplement references}

\usepackage{multirow} 
\usepackage{booktabs} 

\usepackage{color}
\definecolor{deepblue}{rgb}{0,0,0.5}
\definecolor{deepred}{rgb}{0.6,0,0}
\definecolor{deepgreen}{rgb}{0,0.5,0}

\usepackage{listings}
\usepackage[labelformat=simple]{subcaption}

\usepackage{pgfplots}
\pgfplotsset{width=10cm,compat=1.13}

\title{\texttt{mpbn}: a simple tool for efficient edition and analysis of elementary properties of Boolean networks}
\author{
    Van-Giang Trinh\textsuperscript{1}, 
    Belaid Benhamou\textsuperscript{1}, and
Loïc Paulevé\textsuperscript{2}} 

\date{
\begin{small}
\textsuperscript{1}{LIRICA team, LIS, Aix-Marseille University, Marseille 13397, France}\\
\textsuperscript{2}{Univ. Bordeaux, CNRS, Bordeaux INP, LaBRI, UMR5800, F-33400 Talence, France}
\end{small}
}

\begin{document}
\maketitle

\noindent
\begin{small}%
\textbf{Abstract.}
\itshape
The tool \texttt{mpbn} offers a Python programming interface for an easy interactive editing of Boolean networks and the efficient computation of elementary properties of their dynamics, including fixed points, trap spaces, and reachability properties under the Most Permissive update mode.
Relying on Answer-Set Programming logical framework, we show that \texttt{mpbn} is scalable to models with several thousands of nodes and is one of the best-performing tool for computing minimal and maximal
    trap spaces of Boolean networks, a key feature for understanding and controling their stable
    behaviors. The tool is available at \url{https://github.com/bnediction/mpbn}.
\end{small}

\section{Introduction}

Boolean Networks (BNs) are fundamental models of gene regulation and signalling networks dynamics~\cite{glass1973logical,thomas1973boolean,wang2012boolean}, with decades of extensive theoretical and practical research. 
The community developed numerous software tools which implement various algorithms for their analysis~\cite{mussel2010boolnet,chaouiya2012logical,DBLP:journals/bmcsb/HelikarKMBRMWSLR12,klarner2017pyboolnet,naldi2018colomoto,Paulev2020,DBLP:conf/cmsb/TrinhBHS22}, and their edition~\cite{mussel2010boolnet,chaouiya2012logical,DBLP:journals/bmcsb/HelikarKMBRMWSLR12}.
Over the years, BNs have become a versatile formal modeling framework, and many analysis algorithms have been built on elementary properties of BNs, such as fixed points, trap spaces, and dynamical properties~\cite{Paulev2020}.
However, with the constant increase in model size and complexity of Boolean update functions, these previous tools show their performance limitations.
Moreover, most of them do not support interactive and easy-to-use edition and revision of Boolean
network definitions, which, coupled with basic dynamical analyses, would enable seamless experimentation, including for teaching.

Motivated by the above factors, \texttt{mpbn} offers a simple interface for manipulating BNs and for
performing efficient analysis of elementary properties, namely fixed points, trap spaces, as well as
transition graph computation with various update modes, and including most permissive
reachability~\cite{Paulev2020}.

\section{Features}

Formally, a BN is a function from binary vectors of dimension $n$ to themselves, i.e., of the form \(f:\{0,1\}^n \to \{0,1\}^n\). 
These binary vectors are the states (or configurations) of the BN. 
In practice, BNs are specified by $n$ Boolean functions \(f_i : \{0,1\}^n \to \{0,1\}\) for each component \(i \in \{1, \dots, n\}\).
Most of the time, \(f_i\) is expressed as a propositional logic formula, i.e., with Boolean variables \(x_1, \dots, x_n\) and logical connectors.
For example, \(f_2(x) = x_1 \lor \neg (x_3 \land x_4)\), where \(\land\), \(\lor\), \(\neg\) are respectively the conjunction, disjunction, and negation operators.
Also, instead of numbers, components are usually referred to as names (e.g., gene name). 
In that case, we often write a specification like ``\(B := A \lor \neg (C \land D)\)'' instead of the former \(f_2\) expression.

\texttt{mpbn} is an open-source Python module which offers both a simple programming interface, and command line utilities.
We present here its main features.

\subsection{Model edition}

In \texttt{mpbn}, a BN is implemented as a map associating component names to their Boolean functions, expressed in propositional logic. 
It directly supports loading models from the BooleanNet format~\cite{mussel2010boolnet} (\texttt{.bnet}) that is prevalent in the BN research community, and most other usual formats through the \texttt{biolqm} library~\cite{chaouiya2013sbml}.
A model can be created \emph{ab initio} in \texttt{mpbn} and the map can be updated as standard
Python map (\lstinline|dict|) objects using the BooleanNet format.
For example:
\begin{lstlisting}[language=Python]
f = mpbn.MPBooleanNetwork("file.bnet")
f["B"] = "A | !(C & D)"
\end{lstlisting}

The internal representation of functions can be controlled. 
By default, \texttt{mpbn} will convert the expression to its Disjunctive Normal Form (DNF), i.e., disjunction of conjunctions with negation only in front of variables. 
If a variable appears both with and without a negation, \texttt{mpbn} will represent the function using a Binary Decision Diagram (BDD).
\texttt{mpbn} also offers the possibility to preserve the structure as given.
The internal representation can influence the efficiency of the analysis.
Model edited with \texttt{mpbn} can then be exported to text files in the BooleanNet format for analysis with other tools.

\subsection{Most permissive dynamics}

From a state of the BN, an \emph{update mode} enables to compute the next possible states according to $f$.
Traditionally, these update modes reflect the different interleavings and simultaneous applications of component updates: in \emph{synchronous}, all the components are updated at the same time, i.e., there is one single next state being $f(x)$; in \emph{asynchronous}, only one component is updated, leading to several choices for next states, of the form $x_1 \cdots x_{i-1}f_i(x)x_{i+1} \cdots x_n$. 
These dynamics can be summarized by a State Transition Graph (STG), being a directed graph with an edge from state $x$ to $y$ if the update mode enables that transition. 
In \texttt{mpbn} it can be computed using \texttt{f.dynamics("asynchronous")} for instance.
Other classical update modes are also supported, and custom ones can be defined
using \texttt{mpbn} programming interface.

However, it has been shown in~\cite{Paulev2020} that (a)synchronous dynamics can preclude trajectories between states which are actually feasible with quantitative models. 
The recently-introduced \emph{Most Permissive} (MP) update mode overcomes this limitation, by guaranteeing to capture the trajectories of any quantitative model being a \emph{refinement} of the BN, i.e., obtained by introducing threshold, quantities, and kinetics~\cite{Paulev2020}. 
Besides the computation of the full MP dynamics with \texttt{f.dynamics("mp")}, \texttt{mpbn} also implements efficient reachability tests: \texttt{f.reachability(x,y)} returns True if there exists an MP trajectory from state \(x\) to state \(y\). 
Whereas this checking can be rapidly intractable with the (a)synchronous mode, it has been demonstrated that in the MP mode, it can scale to BNs with hundreds of thousands of nodes~\cite{Paulev2020}.
It is worth noting that attractors in the MP mode of a BN (computed by \texttt{f.attractors()}) are identical to minimal trap spaces of this BN~\cite{Paulev2020}.
In addition, we can easily compute reachable MP attractors: \texttt{f.attractors(reachable\_from=x)} returns all the MP attractors reachable from state \(x\).

BNs employing the MP mode (MPBNs) have attracted much attention from researchers in the field of logical modeling of biological systems, such as in \cite{Hrault2023} using MPBNs to unravel new regulatory mechanisms of hematopoietic stem cell aging, and in~\cite{reda2022prioritization} for priorization of candidate genes influencing epilepsies.

\subsection{Fixed points and trap spaces}

A \emph{fixed point} of a BN is a state $x$ such that $f(x) = x$. 
Fixed points are extensively studied in biological modeling as the stable states of the system; their analysis has been a starting point and a standard way for the BN analysis.
A trap space is a particular subcube of $\{0, 1\}^n$~\cite{klarner2015computing,moon2023computational}.
A subcube is characterized by a set of components having a fixed value and can be represented as vectors $c$ in $\{0, 1, \ast\}^n$.
Its vertices are the binary vectors $x = \{0, 1\}^n$ where the components fixed in $c$ have the same value in $x$.
A subcube $c \in \{0, 1, \ast\}^n$ is a \emph{trap space} if, for each of its vertices $x$, $f(x)$ is also one of its vertices.
Intuitively, a trap space represents a ``well-structured'' part of the state space from which it is impossible to escape by applying the functions of the BN.
Notably, trap spaces of a BN are \emph{independent} of its update mode.
An \emph{attractor} is a non-empty set of states such that once entered it, the BN's dynamics cannot escape from it.
In contrast to trap spaces, attractors are \emph{dependent} of the employed update mode of the BN.
Note however that attractor computation is the major analysis on BNs because attractors are linked to biological phenotypes~\cite{thomas1973boolean,wang2012boolean}, and there is a rich history of methods developed for it~\cite{Rozum2021}.

A trap space is \emph{minimal} if it does not contain any other trap space. Fixed points are particular cases of minimal trap spaces where all components are fixed~\cite{klarner2015computing}.  
Minimal trap spaces have been studied as good approximations of attractors of BNs with the asynchronous update mode~\cite{klarner2017pyboolnet,Paulev2020,DBLP:conf/bcb/TrinhHB22}.
A trap space is \emph{maximal} if it is not included in any other trap space, except the full cube \(\{0, 1\}^n\)~\cite{klarner2015computing}.
Maximal trap spaces are also employed for the computation of attractors and for the computation of control strategies to drive the system towards specific attractors~\cite{Rozum2021}, which play a role in systems medicine.
Furthermore, both minimal and maximal trap spaces (also ones restricted to specific subcubes) have been used in several phenotype-based control approaches of biological models~\cite{cifuentes2020control,rozum2021pystablemotifs}.

\texttt{mpbn} implements the enumeration of fixed points, minimal and maximal trap spaces in BNs, possibly restricted to specific subcubes. 
Furthermore, it is possible to perform a partial enumeration of them. 
Under the hood, the computation is performed using logic programming with Answer-Set Programming and the solver \texttt{clingo}~\cite{DBLP:journals/aicom/GebserKKOSS11}.

From extensive benchmarks on real-world and randomly generated models, we observed that \texttt{mpbn} is the sole tool able to address the computation of minimal and maximal trap spaces and fixed points on the full range of models, and with substantially better performance than state-of-the-art tools, especially in large and complex models.
Table~\ref{tab:summary-min} shows a summary of benchmark results for minimal trap spaces on four sets of models with different properties: First, the BBM repository~\cite{pastva2023repository} that consists of 212 published \emph{biologically relevant} BNs, ranging up to 321 variables. 
Second, the set of 23 other non-trivial and \emph{biologically relevant} BNs collected from various bibliographic sources in the literature, ranging up to 4691 variables (see Table 3 of Supplement for more details).
Third, a dataset of Very Large Boolean Networks (VLBN)~\cite{Paulev2020} consisting of 28 \emph{random} BNs, ranging up to 100,000 variables.
And fourth, a dataset (namely AEON) of 100 \emph{random} BNs of up to 1016 variables, generated by using the generator provided in~\cite{DBLP:conf/cav/BenesBPS21}.
In our experiments, we compared \texttt{mpbn} with \texttt{pyboolnet}~\cite{klarner2017pyboolnet} and \texttt{trapmvn}~\cite{trinh2023trap}, which are to our best knowledge the most recent and efficient tools for trap space and fixed point enumeration in BNs.
In addition, we only measured the time to compute the first result, which we believe to make the most fair comparison among the considered techniques because when there are multiple solutions, a benchmark is often testing the technical ability of the implementation to enumerate them quickly, instead of the actual problem-solving.
Note that we here omit the benchmark results for maximal trap spaces and fixed points because they do not reveal any conclusions that are not covered by the minimal trap space case.
All details of the experiments are given in Section ``Experiments'' of Supplement.

\begin{table}
	\caption{Summary of tool performance when computing the first minimal trap space. Columns 2-7 give the number of models completed within the respective time limit.}
	\label{tab:summary-min}
	\centering
    \begin{small}
	\begin{tabular}{c | r r r r r r | }
		\midrule
		\multicolumn{7}{c}{212 BBM models} \\ \midrule
		Method & \texttt{<0.5s} & \texttt{<2s} & \texttt{<10s} & \texttt{<1min} & \texttt{<10min} & \texttt{<1h} \\ \midrule
		\texttt{pyboolnet} & \texttt{141} & \texttt{167} & \texttt{186} & \texttt{190} & \texttt{200} & \texttt{200} \\
		\texttt{trapmvn} & \texttt{206} & \texttt{208} & \texttt{211} & \texttt{211} & \texttt{211} & \texttt{211} \\
		\texttt{mpbn} & \texttt{207} & \texttt{211} & \texttt{212} & \texttt{212} & \texttt{212} & \texttt{212} \\
		\midrule
		\multicolumn{7}{c}{23 selected models} \\ \midrule
		Method & \texttt{<0.5s} & \texttt{<2s} & \texttt{<10s} & \texttt{<1min} & \texttt{<10min} & \texttt{<1h} \\ \midrule
		\texttt{pyboolnet} & \texttt{4} & \texttt{5} & \texttt{5} & \texttt{8} & \texttt{10} & \texttt{10} \\
		\texttt{trapmvn} & \texttt{14} & \texttt{15} & \texttt{16} & \texttt{22} & \texttt{23} & \texttt{23} \\
		\texttt{mpbn} & \texttt{15} & \texttt{20} & \texttt{23} & \texttt{23} & \texttt{23} & \texttt{23} \\
		\midrule
		\multicolumn{7}{c}{28 VLBN models} \\ \midrule
		Method & \texttt{<0.5s} & \texttt{<2s} & \texttt{<10s} & \texttt{<1min} & \texttt{<10min} & \texttt{<1h} \\ \midrule
		\texttt{pyboolnet} & \texttt{0} & \texttt{0} & \texttt{0} & \texttt{0} & \texttt{0} & \texttt{0} \\
		\texttt{trapmvn} & \texttt{0} & \texttt{0} & \texttt{7} & \texttt{8} & \texttt{14} & \texttt{20} \\
		\texttt{mpbn} & \texttt{4} & \texttt{8} & \texttt{16} & \texttt{20} & \texttt{28} & \texttt{28} \\
		\midrule
		\multicolumn{7}{c}{100 AEON models} \\ \midrule
		Method & \texttt{<0.5s} & \texttt{<2s} & \texttt{<10s} & \texttt{<1min} & \texttt{<10min} & \texttt{<1h} \\ \midrule
		\texttt{pyboolnet} & \texttt{0} & \texttt{0} & \texttt{0} & \texttt{0} & \texttt{0} & \texttt{0} \\
		\texttt{trapmvn} & \texttt{0} & \texttt{6} & \texttt{47} & \texttt{96} & \texttt{100} & \texttt{100} \\
		\texttt{mpbn} & \texttt{100} & \texttt{100} & \texttt{100} & \texttt{100} & \texttt{100} & \texttt{100} \\
	\end{tabular}
    \end{small}
\end{table}

\subsection{Example of usage}

\begin{figure}[!htb]
	\includegraphics[height=21.92cm]{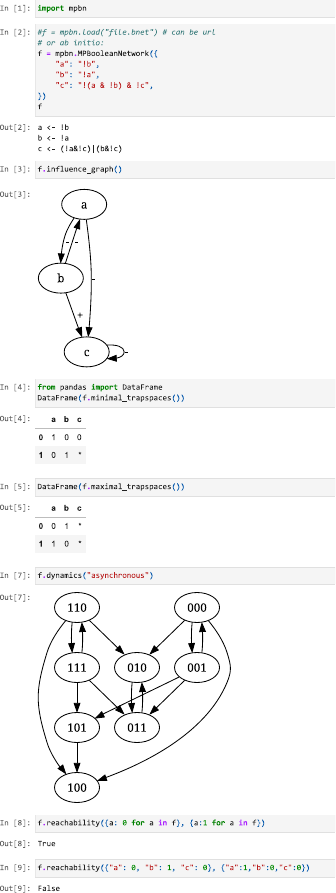}
	\caption{Example of usage within a Jupyter notebook}%
	\label{fig:usage}
\end{figure}

For illustration, Figure~\ref{fig:usage} shows an example of usage for these computations using the Python API. 
The command line utility of \texttt{mpbn} offers similar features.

We consider a BN with three nodes (\(a\), \(b\), and \(c\)) and three associated Boolean functions: \(f_a = \neg b\), \(f_b = \neg a\), and \(f_c = \neg (a \land \neg b) \land \neg c\).
\texttt{mpbn} can load this BN from an input file in the BooleanNet format.
The input file can be located in the local machine or online.
For convenience, \texttt{mpbn} also supports inline initialization, and we use it here (see Row \texttt{In [2]}).
The list of Boolean functions of this BN is printed out in DNF when typing \texttt{f} (see Row \texttt{Out [2]}).
The influence graph of a BN is a signed directed graph where a vertex denotes a node of the BN and an edge \((u, sign, v)\) denotes the effect (activating if \(sign = +\) and inhibiting if \(sign = -\)) of node \(u\) on node \(v\)~\cite{pauleve2012static}.
By typing \texttt{f.influence\_graph()}, the graphical representation of this graph is displayed (see Row \texttt{Out [3]}). 
Then we can compute all minimal and maximal trap spaces of the BN by typing \texttt{f.minimal\_trapspaces()} and \texttt{f.maximal\_trapspaces()}, respectively.
Rows \texttt{Out [4]} and \texttt{Out [5]} show the tabular representations of the results, respectively.
By typing \texttt{f.dynamics("asynchronous")}, the STG of the BN under the asynchronous update mode is obtained and displayed graphically (see Row \texttt{Out [7]}).
Finally, we check if state \texttt{x} reaches state \texttt{y} in the STG by typing \texttt{f.reachability(x, y)}.
For the present STG, \(000\) reaches \(111\) and \(010\) does not reach \(100\).
Hence, the function returns \texttt{True} (see Row \texttt{Out [8]}) and \texttt{False} (see Row \texttt{Out [9]}), respectively.

\section{Conclusion}

\texttt{mpbn} is distributed as a Python PyPi package (\texttt{pip install mpbn}) and conda package (\texttt{conda install -c colomoto -c potassco mpbn}) under the BSD free software license. 
It is also included in the CoLoMoTo Docker distribution~\cite{naldi2018colomoto}. 
Its source code is available at \url{https://github.com/bnediction/mpbn} together with documentation and examples.
Models and scripts used for the benchmarks are provided at \url{https://doi.org/10.5281/zenodo.10781704}.

Combining a simple interface with state-of-the-art resolution algorithms, \texttt{mpbn} is suited for all (1) Boolean model elementary analysis with state of-the-art performance; (2) tool development, as a base (Python) library for implementing advanced analysis, such as attractor analysis and control; and (3) teaching, with interactive edition of BNs, analysis of their basic features, including computation of STGs with different update modes.
We believe that \texttt{mpbn} can become a standard toolbox in the field of Boolean network research
and has the great potential to promote the study of mechanisms and control of biological processes.

\subsubsection*{Acknowledgements}

Work of L.P is supported by the French Agence Nationale pour la Recherche (ANR) in the scope of the project ``BNeDiction'' (grant number ANR-20-CE45-0001).
Work of V-G.T is supported by Institute Carnot STAR, Marseille, France.

\bibliographystyle{unsrt} 
\bibliography{ref}

\newgeometry{margin=3cm}

\onecolumn

\appendix

\begin{center}
    \huge
    Supplement
\end{center}

\bigskip

\lstdefinelanguage{ASP}{%
literate={:-}{$\leftarrow$}1,
morekeywords={not},
numbers=left,
}

\lstset{
language=ASP,
basicstyle=\small\ttfamily,
columns=fullflexible,
keywordstyle=\rm,
firstnumber=last,
morecomment=[l]{\%},
breaklines=true,
breakatwhitespace=true,
keepspaces=true,
commentstyle=\color{gray}\ttfamily,
numbersep=2pt,
numberstyle=\tiny\color{darkgray},
}
\renewcommand{\thelstnumber}{\the\value{lstnumber}}
\lstset{mathescape}

\newcommand{\rev}[1]{#1}

\newcommand{\fmpbn}{\texttt{mpbn} 1.6}
\newcommand{\lmpbn}{\texttt{mpbn} 3.3}

\section{Methods}
\label{sec:Methods}

In the tool \texttt{mpbn}, the computation of fixed points and minimal and maximal trap spaces is performed by the means of Answer-Set Programming (ASP), a declarative logic framework, and the solver
\texttt{clingo}~\citeSupp{DBLP:journals/aicom/GebserKKOSS11}.
Essentially, fixed point and trap space identification problems are translated as logic satisfiability problems, where each solution corresponds to a fixed point or trap space of the input Boolean network.
In this section, we detail a slightly simplified version of the ASP encoding using in \texttt{mpbn},
and which extends the one presented in~\citeSupp{DBLP:conf/ictai/ChevalierFPZ19} by adding support to
non-monotone Boolean functions, i.e., functions that depend both positively and negatively on a same
variable (e.g., exclusive-or functions).

\subsection{Basic definitions}

A \emph{Boolean Network} (BN) is specified by a function $f \colon \mathbb B^n\to\mathbb B^n$, where $\mathbb B = \{0, 1\}$ is the Boolean domain.
A Boolean function $f_i:\mathbb B^n\to\mathbb B$ is \emph{unate} (or \emph{monotone})
if, for each component $j\in\{1,\cdots,n\}$, one of the two following properties holds:
\begin{itemize}
    \item
for every vector $x\in\mathbb B^n$,
$$f_i(x_1,\cdots,x_{j-1},0,x_{j+1},\cdots,x_n) \leq f_i(x_1,\cdots,x_{j-1},1,x_{j+1},\cdots,x_n)$$
    \item
for every vector $x\in\mathbb B^n$,
$$f_i(x_1,\cdots,x_{j-1},0,x_{j+1},\cdots,x_n) \geq f_i(x_1,\cdots,x_{j-1},1,x_{j+1},\cdots,x_n)$$
\end{itemize}
If a Boolean function is not monotone, it is said non-monotone.
E.g., the function ``$x\mapsto (x_1\wedge \neg x_2)\vee (\neg x_1\wedge x_2)$'' is non-monotone, whereas 
$x\mapsto \neg x_1 \vee x_2$ is monotone.
If, for each $i\in\{1,\cdots,n\}$, the (local) function $f_i$ is monotone, then the BN $f$ is said to be
\emph{locally-monotone}; otherwise, the BN $f$ is said non-monotone.

A (sub)\emph{cube} is a set of Boolean vectors which is characterized by a vector in $X \subseteq
\{0,1,\ast\}^n$ and has vertices $\{ x \in\mathbb B^n \mid \forall i\in\{1,\cdots,n\}, X_i \neq
\ast\implies x_i=X_i\}$.
A cube $X$ is a \emph{trap space} for the BN $f$ if it is closed by $f$, i.e., for each vertex $x\in X$ of
the cube, $f(x)$ is also one of its vertices.

In the scope of this supplement,
a \emph{Binary Decision Diagram} (BDD) is a directed acyclic graph with a single root node and two leaf nodes. 
Each leaf node is associated to a value in $\mathbb B$ and each non-leaf node is associated to a BN component in $\{1, \cdots, n\}$.
Each non-leaf node has two children, and every path from the root to a leaf traverses at most one node associated to each BN component.

\subsection{ASP encoding}

We give a very brief overview of Answer-Set Programming (ASP) syntax and semantics that we use in
the next sections; see~\citeSupp{gekakasc12a} for more details.

An ASP program is a Logic Program (LP) being a set
of logical rules with first order logic predicates of the form:
\begin{lstlisting}
$a_0$ :- $a_1$, $\dots$, $a_n$, not $a_{n+1}$, $\dots$, not $a_{n+k}$.
\end{lstlisting}
where $a_i$ are (variable-free) atoms, i.e., elements of the Herbrand base, which is built from all the possible predicates of the LP. 
The Herbrand base is built by instantiating the LP predicates with 
the LP terms (constants or elements of the Herbrand universe).

Essentially, such a logical rule states that when all $a_1,\dots,a_n$ are true  and none of $a_{n+1},\dots,a_{n+k}$ can be proven to be true, then $a_0$ has to be true as well.
Whenever $a_0$ is $\bot$ (false), the rule, also called an \emph{integrity constraint}, becomes:
\begin{lstlisting}
:-$a_1$, $\dots$, $a_n$, not $a_{n+1}$, $\dots$, not $a_{n+k}$.
\end{lstlisting}
Such a rule is satisfied only
if the right hand side of the rule is false (at least one of $a_1,\dots,a_n$ is false or at least
one of $a_{n+1},\dots,a_{n+k}$ is true).
On the other hand,
\lstinline|$a_0$ :- $\top$| ($a_0$ is always true) is abbreviated
as
\lstinline|$a_0$|.

Because our encoding does not involve default negation (\lstinline|not|),  a solution (answer) is a
    subset-minimal Herbrand model (see~\citeSupp{DBLP:journals/ngc/Przymusinski91}), that is, a minimal set of true atoms
where all the logical rules are satisfied.

ASP allows using variables (starting with an upper-case) instead of terms/predicates: these
\emph{template} declarations will be expanded to the corresponding propositional logic rules prior to
the solving, in the \emph{grounding} phase.

We also use the notations
\lstinline|a((x;y))| which is expanded to \lstinline|a(x), a(y)|;
\lstinline|$n$ {a(X): b(X)} $m$|
which is satisfied when at least $n$ and at most $m$ \lstinline|a(X)| are true where \lstinline|X| ranges over
the true \lstinline|b(X)|;
and \lstinline|a(X): b(X)| which is satisfied when for each \lstinline|b(X)| true, \lstinline|a(X)| is
true.
If any term follows such a condition, it is separated with \lstinline|;|.
Finally, rules of form
\begin{lstlisting}
{a} :- $\text{\it body}$.
\end{lstlisting}
leave the choice to make \lstinline|a| true or not whenever the body is satisfied.

\bigskip

In the following, we assume fixing a BN $f \colon \mathbb B^n \to \mathbb B^n$.
\subsubsection{Encoding of Boolean functions}
For each node $a$ of the BN, we declare an atom 
\begin{lstlisting}
node($a$).
\end{lstlisting}
Let $\operatorname{DNF}[f_a] = C_1 \vee \cdots \vee C_k$ be a Disjunctive Normal Form (DNF) of
$f_a$, where $C_i$ is a clause of the form $C_i^1 \wedge \cdots \wedge C_i^{l_i}$ (\(l_i \geq 1\)) where each literal $C_i^m, m \in \{1, \dots, l_i\}$ is
either a node of the BN or the negation of a node.
We require that the DNF is well formed in the sense that no clause subsumes another (i.e., there are no $C_i$, $C_j$, $i\neq j$ such that $\{C_i^1, \cdots, C_i^{l_i}\}\subseteq
\{C_j^1,\cdots,C_j^{l_j}\}$),
and there is no clause containing a literal and its negation.
Whenever $k=0$, i.e., $f_a$ has no clause, $f_a$ is the constant False function
and is declared in our encoding as 
\begin{lstlisting}
constant($a$,-1).
\end{lstlisting}
Whenever $C_1=\emptyset$, i.e., $f_a$ has an empty clause, $f_a$ is the constant True function
and is declared in our encoding as:
\begin{lstlisting}
constant($a$,1).
\end{lstlisting}
Otherwise, for each clause $i\in\{1,\cdots,k\}$ and each $m\in\{1,\cdots,l_i\}$, if the literal $C_i^m$ is
positive, i.e., it is a node $b$, it is declared as:
\begin{lstlisting}
clause($a$,$i$,$b$,1).
\end{lstlisting}
otherwise, i.e., it is the negation of a node $b$, it is declared as:
\begin{lstlisting}
clause($a$,$i$,$b$,-1).
\end{lstlisting}

If there is no node $b$ which appears both positively in one clause $C_i$ and negatively in another clause
$C_j$, $f_a$ is unate (montone). Note that this is only a sufficient condition.
In that case, we add the atom
\begin{lstlisting}
unate($a$).
\end{lstlisting}
Otherwise, \texttt{mpbn} computes a reduced-ordered BDD representation of $f_a$ we denote as $BDD_a$ and encode
its graph representation.
For each non-leaf node $d$ of the BDD testing the BN component $b$ of the BN and having respectively $d_0$ and $d_1$ as low and high children, we declare the following atom:
\begin{lstlisting}
bdd($z$,$b$,($a$,$d_0$),($a$,$d_1$)).
\end{lstlisting}
where $z=a$ if $d$ is the root node, otherwise $z=(a,d)$.
Finally, only the leaf node $d$ for value False is declared:
\begin{lstlisting}
bdd(($z$,$d$),-1).
\end{lstlisting}
where $z$ is defined as above.

It is required that each path from the root to a leaf node of the BDD never traverses two nodes testing a same BN component. This is the case for the BDDs computed in \texttt{mpbn} by employing the \texttt{pyeda} Python library\footnote{\url{https://pyeda.readthedocs.io/en/latest/}}.

\subsubsection{Encoding of Boolean function application}
The following encoding ensures that, for a defined cube
\verb|X|, \lstinline|eval(X,N,-1)| (resp. \lstinline|eval(X,N,1)|) is derived if and only if there
exists a vertex in the cube where the Boolean function of node \verb|N| is False (resp.
True).

The first case is whenever the Boolean function is constant:
\begin{lstlisting}
eval(X,N,V) :- cube(X), node(N), constant(N,V).
\end{lstlisting}

Otherwise, the Boolean function is encoded as DNF using the atoms defined above.
We separated two cases: the evaluation to True, which is always performed using the DNF, and the evaluation to False, which is performed using the DNF representation only whenever the Boolean function is unate.
Otherwise, the evaluation to False is performed by the means of the BDD representation of the Boolean function.

The evaluation from the DNF is performed as follows:
\lstinline|evaldnf(X,N,C,1)| is derived if and only if each literal of the \verb|C|-th clause is
in the cube, i.e., there is one vertex in the cube for which the clause is satisfied.
Similarly, 
\lstinline|evaldnf(X,N,C,-1)| is derived if and only if at least the \emph{negation} of one literal \verb|C|-th clause is
in the cube, i.e., there is one vertex in the cube for which the clause is not satisfied:
\begin{lstlisting}
evaldnf(X,N,C,1) :- cube(X,M,V) : clause(X,M,V); cube(X), clause(N,C,_,_).
evaldnf(X,N,C,-1) :- cube(X,M,-V), clause(N,C,M,V).
\end{lstlisting}

Then, \lstinline|eval(X,N,1)| is derived whenever at least one of its clause has been satisified.
And, whenever the Boolean function is unate,  \lstinline|eval(X,N,-1)| is derived if all the clause
can be unsatisfied:
\begin{lstlisting}
eval(X,N,1) :- evaldnf(X,N,C,1), clause(N,C,_,_).
eval(X,N,-1) :- evaldnf(X,N,C,-1) : clause(N,C,_,_); node(N), cube(X), clause(N,_,_,_); unate(N).
\end{lstlisting}

Finally, whenever the BDD is present, i.e., whenever the Boolean function is not unate,
\lstinline|eval(X,N,-1)| is derived if and only if there exists a path in the BDD which matches with
the cube and which leads to the negative (-1) leaf:
\begin{lstlisting}
eval(X,N,-1) :- evalbdd(X,N,-1), node(N), cube(X).
evalbdd(X,V,V) :- cube(X), V=-1.
evalbdd(X,B,V) :- bdd(B,N,_,HI), cube(X,N,1), evalbdd(X,HI,V).
evalbdd(X,B,V) :- bdd(B,N,LO,_), cube(X,N,-1), evalbdd(X,LO,V).
evalbdd(X,B,V) :- cube(X), bdd(B,V), V=-1.
\end{lstlisting}

\subsubsection{Encoding of fixed points}
We take advantage of the genertic evaluation of Boolean functions over cubes to encode the fixed
point identification problem as follows:
we declare a cube \verb|fb| which consists of a single vertex,
together with an integrity constraint which ensures that the evaluation of each Boolean function on
that cube does not lead to a different value:
\begin{lstlisting}
cube(fp).
1 { cube(fp,N,(-1;1)) } 1 :- node(N).
:- cube(fp,N,V), eval(fp,N,-V).
\end{lstlisting}
The list of fixed points is then obtained by enumerating all the answer sets projected to the atoms
of the form \lstinline|cube(fp,N,V)|.

\subsubsection{Encoding of trap spaces}
Recall that trap spaces are cubes closed by Boolean functions.
Therefore, starting from a first non-empty cube \verb|ts|,
\lstinline|cube(ts,N,V)| is derived whenever \lstinline|eval(X,N,V)| can be derived from the cube.
\begin{lstlisting}
cube(ts).
1 { cube(ts,N,(-1;1)) } :- node(N).
cube(ts,N,V) :- eval(X,N,V).
\end{lstlisting}
Then the set of all trap spaces of the BN one-to-one corresponds to the set of all answer sets projected over the atoms of the form of \lstinline|cube(ts,N,V)|.
From such an anwser set, one can reconstruct the associated trap space $X$ by setting each node
$a$ as free (i.e., $X_a=\ast$) whenver the answer set contains both \lstinline|cube(ts,$a$,-1)| and
\lstinline|cube(ts,$a$,1)|, otherwise, it is fixed to the correspondig value:
$X_a=1$ (resp. $X_a=0$) if only \lstinline|cube(ts,$a$,1| (resp. \lstinline|cube(ts,$a$,-1)|) is in
the answer set.

The minimal and maximal trap spaces are then enumerated by exploiting the domain heuristics of \texttt{clingo} which allows listing only the subset-minimal and subset-maximal answer sets, corresponding exactly and respectively to minimal and maximal trap spaces.

Note that in the case of maximal trap spaces, we need to add a cardinality constraint to exclude from the solution space the trivial trap space $\mathbb B^n$ where $n$ is the number of nodes of the BN:
\begin{lstlisting}
{ cube(ts,N,(-1;1)) : node(N) } $(2n-1)$.
\end{lstlisting}

\textit{Example.}
Let us consider the BN:
	\begin{align*}
		f_1 &= \neg x_2\\
		f_2 &= \neg x_1\\
		f_3 &= \neg (x_1 \land \neg x_2) \land \neg x_3
	\end{align*}
	We have that \(\operatorname{DNF}[f_1] = \neg x_2\), \(\operatorname{DNF}[f_2] = \neg x_1\), and \(\operatorname{DNF}[f_3] = (\neg x_1 \land \neg x_3) \lor (x_2 \land \neg x_3)\).
	All the Boolean functions are unate.
	Then the information about this BN is represented by the set of rules as follows:
\begin{lstlisting}
unate(1).
unate(2).
unate(3).

clause(1,1,2,0).
clause(2,1,1,0).
clause(3,1,1,0).
clause(3,1,3,0).
clause(3,2,2,1).
clause(3,2,3,0).
\end{lstlisting}
	Note that the rules for the Boolean function evaluation are fixed.
	The characterizing rules are fixed for the cases of fixed points and minimal trap spaces.
	For the case of maximal trap spaces, the cardinality constraint is not fixed, it depends on the number of nodes of the BN.
	For the considered BN, it is
\begin{lstlisting}
{ cube(ts,N,(-1;1)) : node(N) } 5.
\end{lstlisting}
	We only consider answer sets projected over the atoms of the form of \lstinline|cube(ts,N,V)|.
	Regarding fixed points, the respective ASP encoding has one answer set \lstinline|{cube(ts, 1, 1), cube(ts, 2, -1), cube(ts, 3, -1)}| that corresponds to the fixed point 100 of the BN.
	Regarding minimal trap spaces, the respective ASP encoding has two subset-minimal answer sets (\lstinline|{cube(ts, 1, -1), cube(ts, 2, 1), cube(ts, 3, -1), cube(ts, 3, 1)}| and \lstinline|{cube(ts, 1, 1), cube(ts, 2, -1), cube(ts, 3, -1)}|) that correspond to two minimal trap spaces (\(01\ast\) and \(100\)) of the BN, respectively.
	Regarding maximal trap spaces, the respective ASP encoding has two subset-maximal answer sets (\lstinline|{cube(ts, 1, 1), cube(ts, 2, -1), cube(ts, 3, -1), cube(ts, 3, 1)}| and \lstinline|{cube(ts, 1, -1), cube(ts, 2, 1), cube(ts, 3, -1), cube(ts, 3, 1)}|) that correspond to two maximal trap spaces (\(10\ast\) and \(01\ast\)) of the BN, respectively.

\section{Experiments}

To evaluate the efficiency of \texttt{mpbn}, we compared it with other state-of-the-art methods regarding three prominent analysis problems on BNs: minimal trap space enumeration, maximal trap space enumeration, and fixed point enumeration.
For each problem instance, we only measured the time to compute the first result, which we believe to make the most fair comparison among the considered methods.
Since we tested a large number of both real-world and random models, we set up a time limit of \emph{one hour} for each method on a problem instance.

All the compared tools are available as Python libraries. 
These libraries along with all benchmark scripts and used data are available at
\url{https://doi.org/10.5281/zenodo.10781704}.
All experiments were performed using a desktop computer running Linux on an Intel(R) Xeon(R) Gold 6244 CPU at 3.60GHz.
We ran the experiments sequentially and in general only used a single core, limited to 32GB of RAM.
The running time was measured internally by each experiment script (i.e., it did not count the overhead of starting the Python interpreter).
However, it included the time for all steps of the computation.

\subsection{Tested tools}

We tested a variety of tools that support the enumeration of
trap spaces and fixed points in BNs.
For each tool, we used its best method for a specific problem type.
Regarding \texttt{mpbn}, we considered two ``milestone'' versions where \texttt{mpbn} 1.6 is the
version described in the original paper~\citeSupp{Paulev2020}, and \lmpbn{} is the latest version that
provides many new methodological materials (see Section~\ref{sec:Methods}), especially the ability
for handling non-monotone BNs, which was not the case for \texttt{mpbn} 1.6.

Table~\ref{tab:summary-tools} summarizes the existing tools of which we are aware and their supported features for BN analysis.
Regarding the minimal trap space enumeration, we omitted \texttt{trappist} because \texttt{trapmvn} is its successor.
\texttt{trapmvn} iterates on \texttt{trappist} by adding heuristics to improve the Petri net encoding process.
We also omitted \texttt{biolqm}~\citeSupp{naldi2018biolqm} that relies on BDDs because it always requires to compute all results, in contrast to other ASP-based methods.
Moreover, due to the inherent drawbacks of both BDDs and the method itself, \texttt{biolqm} is not expected to handle BNs with large size and complex update functions~\citeSupp{Trinh2023tcs}.
With the same reasons, we omitted \texttt{biolqm} for the maximal trap space enumeration and the fixed point enumeration.
Note however that there are three other notable tools that only support the fixed point enumeration: \texttt{an-asp}~\citeSupp{DBLP:journals/almob/AbdallahFRM17}, \texttt{bns}~\citeSupp{DBLP:journals/tcbb/DubrovaT11}, and \texttt{fpcollector}~\citeSupp{DBLP:journals/bioinformatics/AracenaCS21}.
The approaches of \texttt{an-asp} and \texttt{trapmvn} are similar as they rely on the transition-based representations of the original BN, and the constructions both rely on the computation of DNFs of Boolean functions.
In addition, it has been shown in~\citeSupp{trinh2023trap} that \texttt{trapmvn} is comparable to \texttt{an-asp}, even better on average.
\texttt{bns} only supports enumerating all fixed points, and it is limited by the complexity of update functions.
Indeed, it only handles Boolean functions with less than 20 input variables.
When exceeding this bound, it does not compute anything.
Similarly, \texttt{fpcollector} only supports enumerating all fixed points and it is indeed not expected to handle large and complex BNs.
Hence, we omitted \texttt{an-asp}, \texttt{bns}, and \texttt{fpcollector} for the fixed point enumeration.
Overall, we have four tested tools for each of the three problem types: \texttt{mpbn} 1.6, \lmpbn{}, \texttt{pyboolnet}, and \texttt{trapmvn}.
It is worth noting that to our best knowledge \texttt{pyboolnet} and \texttt{trapmvn} are the most recent and efficient methods for such kinds of analysis on BNs, which is justified by the extensive experiments conducted by~\citeSupp{klarner2015computing,Trinh2023tcs,trinh2023trap}.

\begin{table}[!ht]
	\caption{Summary of existing tools and their supported features for BN analysis. Columns 2-4 indicate the minimal trap space enumeration, the maximal trap space enumeration, and the fixed point enumeration, respectively.}
	\label{tab:summary-tools}
	\centering
	\begin{tabular}{l|m{2cm}m{2cm}m{2cm}}
		\midrule
		Tool & Minimal & Maximal & Fixed point \\ \midrule
		\texttt{mpbn} & yes & yes & yes \\
		\texttt{pyboolnet}~\citeSupp{klarner2017pyboolnet} & yes & yes & yes \\
		\texttt{trappist}~\citeSupp{DBLP:conf/cmsb/TrinhBHS22,Trinh2023tcs} & yes & no & no \\
		\texttt{trapmvn}~\citeSupp{trinh2023trap} & yes & yes & yes \\
		\texttt{biolqm}~\citeSupp{naldi2018biolqm} & yes & yes & yes \\
		\texttt{an-asp}~\citeSupp{DBLP:journals/almob/AbdallahFRM17} & no & no & yes \\
		\texttt{bns}~\citeSupp{DBLP:journals/tcbb/DubrovaT11} & no & no & yes \\
		\texttt{fpcollector}~\citeSupp{DBLP:journals/bioinformatics/AracenaCS21} & no & no & yes \\
	\end{tabular}
\end{table}

\subsection{Tested models}

We tested the tools on a wide range of real-world and random models. In particular, we used:

\paragraph{Biodivine Boolean Models (BBM)} The BBM repository~\citeSupp{pastva2023repository} provides a wide variety of published \emph{biologically relevant} BN models. Here, we used the August 2022 edition\footnote{\url{https://github.com/sybila/biodivine-boolean-models/releases/tag/august-2022}} that consists of 212 models, ranging up to 321 variables and 1100 regulations.

\paragraph{Manually selected models} To supplement the BBM dataset, we also performed a separate survey of related literature and found 23 additional \emph{biologically relevant} models.
These models are non-trivial, even contain a very large number of nodes.
A summary of these models, ranging up to 4691 variables, is given in Table~\ref{tab:selected-real}.

\paragraph{Very Large Boolean Networks (VLBN)} The VLBN dataset\footnote{\url{https://doi.org/10.5281/zenodo.3714875}} provides 28 \emph{random} BNs with scale-free topology and inhibitor-dominant update functions~\citeSupp{Paulev2020}. 
The models range up to 100.000 variables. 
They are very large, but have rather simple non-monotone update functions.

\paragraph{Manually generated models} To supplement the VLBN dataset, we also used a dataset (namely AEON) of 100 \emph{random} BNs generated by the generator of~\citeSupp{DBLP:conf/cav/BenesBPS21}, ranging up to 1016 variables. 
This generator uses a degree distribution based on the BBM dataset to sample the network topology. 
For update functions, it samples from a subset of nested-canalizing unate functions.
Compared to the VLBN dataset, it thus covers a smaller range of sizes, but a wider range of update functions.

\begin{table}[!ht]
	\centering%
	\caption{Selected real-world models. The table lists total variable count $n$, the number of source (i.e., input) variables $s$, and the respective bibliographic reference.}
	\label{tab:selected-real}
	\begin{tabular}{clrrl}
		\toprule
		\textbf{No.} & \textbf{Filename} & \textbf{\(n\)} & \textbf{\(s\)} & \textbf{Source} \\
		\midrule
		\rownb & \texttt{Rho-family\_GTPases\_signaling} & 33 & 1 & \citeSupp{DesignPrinciplesGeneNetworks} \\
		\rownb & \texttt{tca\_cycle} & 69 & 35 & \citeSupp{hemedan2023applications}\\
		\rownb & \texttt{Executable\_...\_mast\_cell\_...\_BCC} & 73 & 19 & \citeSupp{aghamiri2020automated} \\
		\rownb & \texttt{Leukaemia.free-inputs} & 107 & 6 & \citeSupp{trinh2023trap} \\
		\rownb & \texttt{EMT\_Mechanosensing} & 136 & 4 & \citeSupp{Sullivan2022} \\
		\rownb & \texttt{angiofull} & 142 & 28 & \citeSupp{Weinstein2017} \\
		\rownb & \texttt{EMT\_Mechanosensing\_TGFbeta} & 150 & 6 & \citeSupp{Sullivan2022} \\
		\rownb & \texttt{VPC.free-inputs} & 169 & 7 & \citeSupp{trinh2023trap} \\
		\rownb & \texttt{InVivo.free-inputs} & 179 & 7 & \citeSupp{trinh2023trap} \\
		\rownb & \texttt{Executable\_...\_MAPK\_model\_BCC} & 181 & 37 & \citeSupp{aghamiri2020automated} \\
		\rownb & \texttt{InVitro.free-inputs} & 185 & 9 & \citeSupp{trinh2023trap} \\
		\rownb & \texttt{T-cell\_co-receptor\_molecules\_calcium\_channel} & 206 & 39 & \citeSupp{Ganguli2015} \\
		\rownb & \texttt{SkinModel.free-inputs} & 300 & 10 & \citeSupp{trinh2023trap} \\
		\rownb & \texttt{Leishmania} & 342 & 81 & \citeSupp{DBLP:journals/ejbsb/GanguliCCS15} \\
		\rownb & \texttt{Metabolism\_demo.free-inputs} & 355 & 9 & \citeSupp{trinh2023trap} \\
		\rownb & \texttt{Executable\_file\_for\_cholocystokinin\_model\_BCC} & 383 & 74 & \citeSupp{aghamiri2020automated} \\
		\rownb & \texttt{ra\_map} & 447 & 125 & \citeSupp{Singh2023} \\
		\rownb & \texttt{CAF-model} & 463 & 62 & \citeSupp{Aghakhani2023.05.11.540378} \\
		\rownb & \texttt{Executable\_file\_for\_Alzheimer\_model\_BCC} & 762 & 237 & \citeSupp{aghamiri2020automated} \\
		\rownb & \texttt{S1\_Table} & 1659 & 521 & \citeSupp{DBLP:conf/cmsb/TrinhBHS22} \\
		\rownb & \texttt{Human\_network} & 1953 & 669 & \citeSupp{DBLP:conf/cmsb/TrinhBHS22} \\
		\rownb & \texttt{SN5} & 2746 & 829 & \citeSupp{DBLP:conf/cmsb/TrinhBHS22} \\
		\rownb & \texttt{turei\_2016} & 4691 & 1257 & \citeSupp{DBLP:conf/cmsb/TrinhBHS22} \\
		\bottomrule
	\end{tabular}
\end{table}

\subsection{Results on minimal trap spaces}

Regarding the minimal trap space enumeration, Table~\ref{tab:summary-min} summarizes the obtained results.
We here reported the number of models completed within a specific time limit for each tool on each dataset.
To better visualize the performance scaling of individual tools on the real-world and random models, we prepared Figures~\ref{fig:cumulative-real-min} and~\ref{fig:cumulative-random-min}, respectively. 
These figures show in detail how many models can be completed by each tool within a specific time limit.
There are 235 models come from the BBM repository and the selected models and 128 random models come from the VLBN and AEON datasets.

\begin{table}[!ht]
	\caption{Summary of tool performance when computing the first minimal trap space. Columns 2-7 give the number of models completed within the respective time limit.}
	\label{tab:summary-min}
	\centering
	\begin{tabular}{c | r r r r r r | }
		\midrule
		\multicolumn{7}{c}{212 BBM models} \\ \midrule
		Method & \texttt{<0.5s} & \texttt{<2s} & \texttt{<10s} & \texttt{<1min} & \texttt{<10min} & \texttt{<1h} \\ \midrule
		\texttt{pyboolnet} & \texttt{141} & \texttt{166} & \texttt{187} & \texttt{190} & \texttt{200} & \texttt{200} \\
		\texttt{trapmvn} & \texttt{206} & \texttt{208} & \texttt{211} & \texttt{211} & \texttt{211} & \texttt{211} \\
		\fmpbn{} & \texttt{99} & \texttt{99} & \texttt{99} & \texttt{99} & \texttt{99} & \texttt{99} \\
		\lmpbn{} & \texttt{207} & \texttt{211} & \texttt{212} & \texttt{212} & \texttt{212} & \texttt{212} \\
		\midrule
		\multicolumn{7}{c}{23 selected models} \\ \midrule
		Method & \texttt{<0.5s} & \texttt{<2s} & \texttt{<10s} & \texttt{<1min} & \texttt{<10min} & \texttt{<1h} \\ \midrule
		\texttt{pyboolnet} & \texttt{4} & \texttt{5} & \texttt{5} & \texttt{8} & \texttt{12} & \texttt{14} \\
		\texttt{trapmvn} & \texttt{14} & \texttt{15} & \texttt{16} & \texttt{22} & \texttt{23} & \texttt{23} \\
		\fmpbn{} & \texttt{5} & \texttt{8} & \texttt{9} & \texttt{9} & \texttt{9} & \texttt{9} \\
		\lmpbn{} & \texttt{15} & \texttt{20} & \texttt{23} & \texttt{23} & \texttt{23} & \texttt{23} \\
		\midrule
		\multicolumn{7}{c}{28 VLBN models} \\ \midrule
		Method & \texttt{<0.5s} & \texttt{<2s} & \texttt{<10s} & \texttt{<1min} & \texttt{<10min} & \texttt{<1h} \\ \midrule
		\texttt{pyboolnet} & \texttt{0} & \texttt{0} & \texttt{0} & \texttt{0} & \texttt{0} & \texttt{0} \\
		\texttt{trapmvn} & \texttt{0} & \texttt{0} & \texttt{7} & \texttt{8} & \texttt{14} & \texttt{20} \\
		\fmpbn{} & \texttt{1} & \texttt{7} & \texttt{15} & \texttt{20} & \texttt{28} & \texttt{28} \\
		\lmpbn{} & \texttt{4} & \texttt{8} & \texttt{16} & \texttt{20} & \texttt{28} & \texttt{28} \\
		\midrule
		\multicolumn{7}{c}{100 AEON models} \\ \midrule
		Method & \texttt{<0.5s} & \texttt{<2s} & \texttt{<10s} & \texttt{<1min} & \texttt{<10min} & \texttt{<1h} \\ \midrule
		\texttt{pyboolnet} & \texttt{0} & \texttt{0} & \texttt{0} & \texttt{0} & \texttt{0} & \texttt{0} \\
		\texttt{trapmvn} & \texttt{0} & \texttt{6} & \texttt{47} & \texttt{96} & \texttt{100} & \texttt{100} \\
		\fmpbn{} & \texttt{100} & \texttt{100} & \texttt{100} & \texttt{100} & \texttt{100} & \texttt{100} \\
		\lmpbn{} & \texttt{100} & \texttt{100} & \texttt{100} & \texttt{100} & \texttt{100} & \texttt{100} \\
	\end{tabular}
\end{table}

\begin{figure}[!ht]
	\centering
	\definecolor{color-pyboolnet}{HTML}{1b5e20}
	\definecolor{color-trapmvn}{HTML}{0376BB}			
	\definecolor{color-mpbn-3.2}{HTML}{e65100}	
	\definecolor{color-mpbn-1.6}{HTML}{69c1a4}	
	\pgfplotsset{width=8cm,height=7cm}
	\centering
	\begin{tikzpicture}
	\begin{axis}[
	xlabel = {Runtime (log)},
	ylabel = {Completed instances},
	log ticks with fixed point,
	xtick={0.5,2,10,60,600,3600},
	xticklabels={0.5s,2s,10s,1min,10min,1h},
	ytick={99, 150, 200, 250},
	xmin = 0.3,
	xmax = 3600,
	ymin = 90,
	ymax = 250,
	xmode=log,
	grid=both,
	grid style={line width=.1pt, draw=gray!10},
	major grid style={line width=.2pt,draw=gray!50},
	]
	\addplot [color-pyboolnet, line width=1.25pt] table {results/min-aggregated/real-pyboolnet.tsv};		
	\addplot [color-trapmvn, line width=1.25pt] table {results/min-aggregated/real-trapmvn.tsv};
	\addplot [color-mpbn-1.6, line width=1.25pt] table {results/min-aggregated/real-mpbn-1.6.tsv};
	\addplot [color-mpbn-3.2, line width=1.25pt] table {results/min-aggregated/real-mpbn-3.2.tsv};													
	\end{axis}
	\end{tikzpicture}
	
	\texttt{pyboolnet} {\color{color-pyboolnet} $\blacksquare$} / \texttt{trapmvn} {\color{color-trapmvn} $\blacksquare$} / \fmpbn{} {\color{color-mpbn-1.6} $\blacksquare$} / \lmpbn{} {\color{color-mpbn-3.2} $\blacksquare$}
	\caption{Cumulative minimal trap space experiments completed (y-axis) until a specific time point (x-axis, logarithmic). Concerns the 235 real-world models.}
	\label{fig:cumulative-real-min}
\end{figure}

\begin{figure}[!ht]
	\centering
	\definecolor{color-pyboolnet}{HTML}{1b5e20}
	\definecolor{color-trapmvn}{HTML}{0376BB}			
	\definecolor{color-mpbn-3.2}{HTML}{e65100}	
	\definecolor{color-mpbn-1.6}{HTML}{69c1a4}	
	\pgfplotsset{width=8cm,height=7cm}
	\centering
	\begin{tikzpicture}
	\begin{axis}[
	xlabel = {Runtime (log)},
	ylabel = {Completed instances},
	log ticks with fixed point,
	xtick={0.5,2,10,60,600,3600},
	xticklabels={0.5s,2s,10s,1min,10min,1h},
	ytick={0, 25, 50, 75, 100, 130},
	xmin = 0.3,
	xmax = 3600,
	ymin = 0,
	ymax = 130,
	xmode=log,
	grid=both,
	grid style={line width=.1pt, draw=gray!10},
	major grid style={line width=.2pt,draw=gray!50},
	]
	\addplot [color-pyboolnet, line width=1.25pt] table {results/min-aggregated/random-pyboolnet.tsv};		
	\addplot [color-trapmvn, line width=1.25pt] table {results/min-aggregated/random-trapmvn.tsv};
	\addplot [color-mpbn-1.6, line width=1.25pt] table {results/min-aggregated/random-mpbn-1.6.tsv};
	\addplot [color-mpbn-3.2, line width=1.25pt] table {results/min-aggregated/random-mpbn-3.2.tsv};													
	\end{axis}
	\end{tikzpicture}
	
	\texttt{pyboolnet} {\color{color-pyboolnet} $\blacksquare$} / \texttt{trapmvn} {\color{color-trapmvn} $\blacksquare$} / \fmpbn{} {\color{color-mpbn-1.6} $\blacksquare$} / \lmpbn{} {\color{color-mpbn-3.2} $\blacksquare$}
	\caption{Cumulative minimal trap space experiments completed (y-axis) until a specific time point (x-axis, logarithmic). Concerns the 128 randomly generated models.}
	\label{fig:cumulative-random-min}
\end{figure}

For real-world models, we first observed that \lmpbn{} completely overcomes the limitation on
non-monotone BNs of \fmpbn{}.
\lmpbn{} completed much more models than \fmpbn{} for every time limit and all such extra models are
non-monotone.
This result confirms the efficacy of the new methedological materials that we proposed and implemented in \lmpbn{}.
Second, we can see that \lmpbn{} is the fastest tool as it completed more models than all the other tools for every time limit.
In particular, it is the sole tool that can handle all the BBM models with the time limit 1h.
Actually, it completed every BBM model (also selected model) within 10s.
Third, although \lmpbn{} is better than \texttt{pyboolnet} and \texttt{trapmvn}, their performance differences are not much exhibited.
This can be explained by the fact that in most of the real-world models, update functions have quite simple forms, even sometimes just simple conjunctions or disjunctions of literals.

For random models, the performance difference between \fmpbn{} and \lmpbn{} is negligible because
all the considered random models are locally-monotone.
Second, \texttt{pyboolnet} could not handle any random model.
It is not surprising because these models have complex update functions, which is the main weakness of the ASP encoding based on prime implicants~\citeSupp{Trinh2023tcs}.
Third, \texttt{trapmvn} completed all the AEON models with the time limit of 10min, but it completed only 20 VLBN models (it failed with all 50.000-node and 100.000-node models) with the time limit of 1h.
The reason is that for random models, the number of transitions of the Petri net encoding is large, making the ASP encoding more complicated.
Moreover, for the majority of the succeeded models, it took more than 10s each.
Finally, \lmpbn{} vastly outperforms \texttt{trapmvn}.
It completed much more models than \texttt{trapmvn} for every time limit in the case of VLBN models and for all time limits of at most 10s in the case of AEON models.
In particular, it is the sole tool that can handle all VLBN models with the time limit of 1h (actually it can handle every such model within 10min) and it can handle every AEON model within only 0.5s.

\subsection{Results on maximal trap spaces}

Regarding the maximal trap space enumeration, the results are presented in Table~\ref{tab:summary-max} and Figures~\ref{fig:cumulative-real-max} and~\ref{fig:cumulative-random-max}.
These results do not reveal any conclusions that are not covered by the minimal trap space case.

\begin{table}[!ht]
	\caption{Summary of tool performance when computing the first maximal trap space. Columns 2-7 give the number of models completed within the respective time limit.}
	\label{tab:summary-max}
	\centering
	\begin{tabular}{c | r r r r r r | }
		\midrule
		\multicolumn{7}{c}{212 BBM models} \\ \midrule
		Method & \texttt{<0.5s} & \texttt{<2s} & \texttt{<10s} & \texttt{<1min} & \texttt{<10min} & \texttt{<1h} \\ \midrule
		\texttt{pyboolnet} & \texttt{140} & \texttt{166} & \texttt{188} & \texttt{190} & \texttt{200} & \texttt{200} \\
		\texttt{trapmvn} & \texttt{206} & \texttt{208} & \texttt{211} & \texttt{211} & \texttt{211} & \texttt{211} \\
		\fmpbn{} & \texttt{99} & \texttt{99} & \texttt{99} & \texttt{99} & \texttt{99} & \texttt{99} \\
		\lmpbn{} & \texttt{207} & \texttt{211} & \texttt{212} & \texttt{212} & \texttt{212} & \texttt{212} \\
		\midrule
		\multicolumn{7}{c}{23 selected models} \\ \midrule
		Method & \texttt{<0.5s} & \texttt{<2s} & \texttt{<10s} & \texttt{<1min} & \texttt{<10min} & \texttt{<1h} \\ \midrule
		\texttt{pyboolnet} & \texttt{4} & \texttt{5} & \texttt{5} & \texttt{8} & \texttt{12} & \texttt{14} \\
		\texttt{trapmvn} & \texttt{14} & \texttt{15} & \texttt{16} & \texttt{22} & \texttt{23} & \texttt{23} \\
		\fmpbn{} & \texttt{5} & \texttt{8} & \texttt{9} & \texttt{9} & \texttt{9} & \texttt{9} \\
		\lmpbn{} & \texttt{15} & \texttt{19} & \texttt{23} & \texttt{23} & \texttt{23} & \texttt{23} \\
		\midrule
		\multicolumn{7}{c}{28 VLBN models} \\ \midrule
		Method & \texttt{<0.5s} & \texttt{<2s} & \texttt{<10s} & \texttt{<1min} & \texttt{<10min} & \texttt{<1h} \\ \midrule
		\texttt{pyboolnet} & \texttt{0} & \texttt{0} & \texttt{0} & \texttt{0} & \texttt{0} & \texttt{0} \\
		\texttt{trapmvn} & \texttt{0} & \texttt{0} & \texttt{7} & \texttt{10} & \texttt{16} & \texttt{16} \\
		\fmpbn{} & \texttt{1} & \texttt{8} & \texttt{15} & \texttt{20} & \texttt{28} & \texttt{28} \\
		\lmpbn{} & \texttt{4} & \texttt{8} & \texttt{16} & \texttt{23} & \texttt{28} & \texttt{28} \\
		\midrule
		\multicolumn{7}{c}{100 AEON models} \\ \midrule
		Method & \texttt{<0.5s} & \texttt{<2s} & \texttt{<10s} & \texttt{<1min} & \texttt{<10min} & \texttt{<1h} \\ \midrule
		\texttt{pyboolnet} & \texttt{0} & \texttt{0} & \texttt{0} & \texttt{0} & \texttt{0} & \texttt{0} \\
		\texttt{trapmvn} & \texttt{0} & \texttt{7} & \texttt{52} & \texttt{96} & \texttt{100} & \texttt{100} \\
		\fmpbn{} & \texttt{100} & \texttt{100} & \texttt{100} & \texttt{100} & \texttt{100} & \texttt{100} \\
		\lmpbn{} & \texttt{100} & \texttt{100} & \texttt{100} & \texttt{100} & \texttt{100} & \texttt{100} \\
	\end{tabular}
\end{table}

\begin{figure}[!ht]
	\centering
	\definecolor{color-pyboolnet}{HTML}{1b5e20}
	\definecolor{color-trapmvn}{HTML}{0376BB}			
	\definecolor{color-mpbn-3.2}{HTML}{e65100}	
	\definecolor{color-mpbn-1.6}{HTML}{69c1a4}	
	\pgfplotsset{width=8cm,height=7cm}
	\centering
	\begin{tikzpicture}
	\begin{axis}[
	xlabel = {Runtime (log)},
	ylabel = {Completed instances},
	log ticks with fixed point,
	xtick={0.5,2,10,60,600, 3600},
	xticklabels={0.5s,2s,10s,1min,10min,1h},
	ytick={99, 150, 200, 250},
	xmin = 0.3,
	xmax = 3600,
	ymin = 90,
	ymax = 250,
	xmode=log,
	grid=both,
	grid style={line width=.1pt, draw=gray!10},
	major grid style={line width=.2pt,draw=gray!50},
	]
	\addplot [color-pyboolnet, line width=1.25pt] table {results/max-aggregated/real-pyboolnet.tsv};		
	\addplot [color-trapmvn, line width=1.25pt] table {results/max-aggregated/real-trapmvn.tsv};
	\addplot [color-mpbn-1.6, line width=1.25pt] table {results/max-aggregated/real-mpbn-1.6.tsv};
	\addplot [color-mpbn-3.2, line width=1.25pt] table {results/max-aggregated/real-mpbn-3.2.tsv};													
	\end{axis}
	\end{tikzpicture}
	
	\texttt{pyboolnet} {\color{color-pyboolnet} $\blacksquare$} / \texttt{trapmvn} {\color{color-trapmvn} $\blacksquare$} / \fmpbn{} {\color{color-mpbn-1.6} $\blacksquare$} / \lmpbn{} {\color{color-mpbn-3.2} $\blacksquare$}
	\caption{Cumulative maximal trap space experiments completed (y-axis) until a specific time point (x-axis, logarithmic). Concerns the 235 real-world models.}
	\label{fig:cumulative-real-max}
\end{figure}

\begin{figure}[!ht]
	\centering
	\definecolor{color-pyboolnet}{HTML}{1b5e20}
	\definecolor{color-trapmvn}{HTML}{0376BB}			
	\definecolor{color-mpbn-3.2}{HTML}{e65100}	
	\definecolor{color-mpbn-1.6}{HTML}{69c1a4}	
	\pgfplotsset{width=8cm,height=7cm}
	\centering
	\begin{tikzpicture}
	\begin{axis}[
	xlabel = {Runtime (log)},
	ylabel = {Completed instances},
	log ticks with fixed point,
	xtick={0.5,2,10,60,600,3600},
	xticklabels={0.5s,2s,10s,1min,10min,1h},
	ytick={0, 25, 50, 75, 100, 130},
	xmin = 0.3,
	xmax = 3600,
	ymin = 0,
	ymax = 130,
	xmode=log,
	grid=both,
	grid style={line width=.1pt, draw=gray!10},
	major grid style={line width=.2pt,draw=gray!50},
	]
	\addplot [color-pyboolnet, line width=1.25pt] table {results/max-aggregated/random-pyboolnet.tsv};		
	\addplot [color-trapmvn, line width=1.25pt] table {results/max-aggregated/random-trapmvn.tsv};
	\addplot [color-mpbn-1.6, line width=1.25pt] table {results/max-aggregated/random-mpbn-1.6.tsv};
	\addplot [color-mpbn-3.2, line width=1.25pt] table {results/max-aggregated/random-mpbn-3.2.tsv};													
	\end{axis}
	\end{tikzpicture}
	
	\texttt{pyboolnet} {\color{color-pyboolnet} $\blacksquare$} / \texttt{trapmvn} {\color{color-trapmvn} $\blacksquare$} / \fmpbn{} {\color{color-mpbn-1.6} $\blacksquare$} / \lmpbn{} {\color{color-mpbn-3.2} $\blacksquare$}
	\caption{Cumulative maximal trap space experiments completed (y-axis) until a specific time point (x-axis, logarithmic). Concerns the 128 randomly generated models.}
	\label{fig:cumulative-random-max}
\end{figure}

\subsection{Results on fixed points}

\begin{figure}[!ht]
	\centering
	\definecolor{color-pyboolnet}{HTML}{1b5e20}
	\definecolor{color-trapmvn}{HTML}{0376BB}			
	\definecolor{color-mpbn-3.2}{HTML}{e65100}	
	\definecolor{color-mpbn-1.6}{HTML}{69c1a4}	
	\pgfplotsset{width=8cm,height=7cm}
	\centering
	\begin{tikzpicture}
	\begin{axis}[
	xlabel = {Runtime (log)},
	ylabel = {Completed instances},
	log ticks with fixed point,
	xtick={0.5,2,10,60,600,3600},
	xticklabels={0.5s,2s,10s,1min,10min,1h},
	ytick={99, 150, 200, 250},
	xmin = 0.3,
	xmax = 3600,
	ymin = 90,
	ymax = 250,
	xmode=log,
	grid=both,
	grid style={line width=.1pt, draw=gray!10},
	major grid style={line width=.2pt,draw=gray!50},
	]
	\addplot [color-pyboolnet, line width=1.25pt] table {results/fix-aggregated/real-pyboolnet.tsv};		
	\addplot [color-trapmvn, line width=1.25pt] table {results/fix-aggregated/real-trapmvn.tsv};
	\addplot [color-mpbn-1.6, line width=1.25pt] table {results/fix-aggregated/real-mpbn-1.6.tsv};
	\addplot [color-mpbn-3.2, line width=1.25pt] table {results/fix-aggregated/real-mpbn-3.2.tsv};													
	\end{axis}
	\end{tikzpicture}
	
	\texttt{pyboolnet} {\color{color-pyboolnet} $\blacksquare$} / \texttt{trapmvn} {\color{color-trapmvn} $\blacksquare$} / \fmpbn{} {\color{color-mpbn-1.6} $\blacksquare$} / \lmpbn{} {\color{color-mpbn-3.2} $\blacksquare$}
	\caption{Cumulative fixed point experiments completed (y-axis) until a specific time point (x-axis, logarithmic). Concerns the 235 real-world models.}
	\label{fig:cumulative-real-fix}
\end{figure}
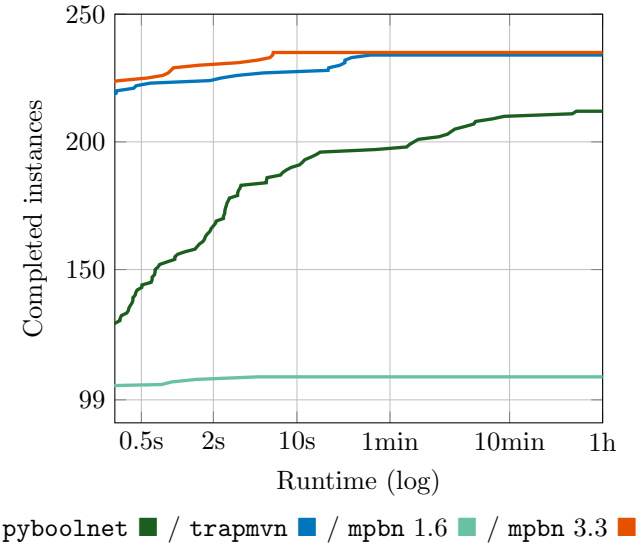

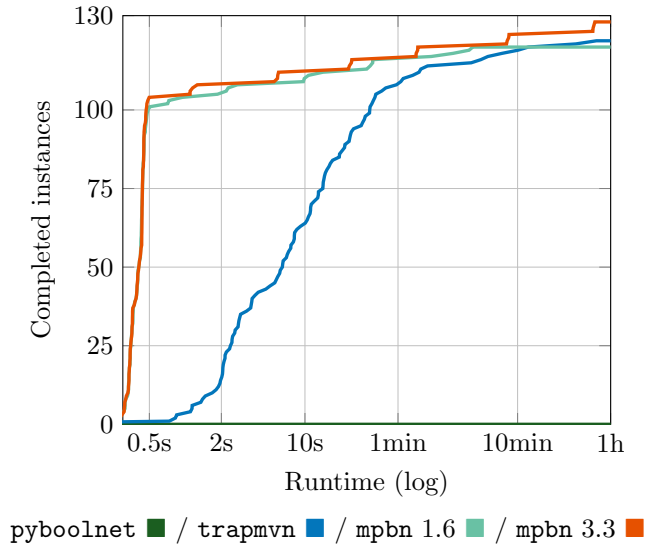
\begin{figure}[!ht]
	\centering
	\definecolor{color-pyboolnet}{HTML}{1b5e20}
	\definecolor{color-trapmvn}{HTML}{0376BB}			
	\definecolor{color-mpbn-3.2}{HTML}{e65100}	
	\definecolor{color-mpbn-1.6}{HTML}{69c1a4}	
	\pgfplotsset{width=8cm,height=7cm}
	\centering
	\begin{tikzpicture}
	\begin{axis}[
	xlabel = {Runtime (log)},
	ylabel = {Completed instances},
	log ticks with fixed point,
	xtick={0.5,2,10,60,600,3600},
	xticklabels={0.5s,2s,10s,1min,10min,1h},
	ytick={0, 25, 50, 75, 100, 130},
	xmin = 0.3,
	xmax = 3600,
	ymin = 0,
	ymax = 130,
	xmode=log,
	grid=both,
	grid style={line width=.1pt, draw=gray!10},
	major grid style={line width=.2pt,draw=gray!50},
	]
	\addplot [color-pyboolnet, line width=1.25pt] table {results/fix-aggregated/random-pyboolnet.tsv};		
	\addplot [color-trapmvn, line width=1.25pt] table {results/fix-aggregated/random-trapmvn.tsv};
	\addplot [color-mpbn-1.6, line width=1.25pt] table {results/fix-aggregated/random-mpbn-1.6.tsv};
	\addplot [color-mpbn-3.2, line width=1.25pt] table {results/fix-aggregated/random-mpbn-3.2.tsv};													
	\end{axis}
	\end{tikzpicture}
	
	\texttt{pyboolnet} {\color{color-pyboolnet} $\blacksquare$} / \texttt{trapmvn} {\color{color-trapmvn} $\blacksquare$} / \fmpbn{} {\color{color-mpbn-1.6} $\blacksquare$} / \lmpbn{} {\color{color-mpbn-3.2} $\blacksquare$}
	\caption{Cumulative fixed point experiments completed (y-axis) until a specific time point (x-axis, logarithmic). Concerns the 128 randomly generated models.}
	\label{fig:cumulative-random-fix}
\end{figure}

Regarding the fixed point enumeration, the results are presented in Table~\ref{tab:summary-fix} and Figures~\ref{fig:cumulative-real-fix} and~\ref{fig:cumulative-random-fix}.
These results do not reveal any conclusions that are not covered by the minimal trap space case.

\begin{table}[!ht]
	\caption{Summary of tool performance when computing the first fixed point. Columns 2-7 give the number of models completed within the respective time limit.}
	\label{tab:summary-fix}
	\centering
	\begin{tabular}{c | r r r r r r | }
		\midrule
		\multicolumn{7}{c}{212 BBM models} \\ \midrule
		Method & \texttt{<0.5s} & \texttt{<2s} & \texttt{<10s} & \texttt{<1min} & \texttt{<10min} & \texttt{<1h} \\ \midrule
		\texttt{pyboolnet} & \texttt{139} & \texttt{162} & \texttt{185} & \texttt{189} & \texttt{199} & \texttt{199} \\
		\texttt{trapmvn} & \texttt{207} & \texttt{209} & \texttt{211} & \texttt{211} & \texttt{211} & \texttt{211} \\
		\fmpbn{} & \texttt{99} & \texttt{99} & \texttt{99} & \texttt{99} & \texttt{99} & \texttt{99} \\
		\lmpbn{} & \texttt{209} & \texttt{211} & \texttt{212} & \texttt{212} & \texttt{212} & \texttt{212} \\
		\midrule
		\multicolumn{7}{c}{23 selected models} \\ \midrule
		Method & \texttt{<0.5s} & \texttt{<2s} & \texttt{<10s} & \texttt{<1min} & \texttt{<10min} & \texttt{<1h} \\ \midrule
		\texttt{pyboolnet} & \texttt{3} & \texttt{5} & \texttt{5} & \texttt{8} & \texttt{11} & \texttt{13} \\
		\texttt{trapmvn} & \texttt{15} & \texttt{15} & \texttt{16} & \texttt{23} & \texttt{23} & \texttt{23} \\
		\fmpbn{} & \texttt{5} & \texttt{8} & \texttt{9} & \texttt{9} & \texttt{9} & \texttt{9} \\
		\lmpbn{} & \texttt{15} & \texttt{19} & \texttt{23} & \texttt{23} & \texttt{23} & \texttt{23} \\
		\midrule
		\multicolumn{7}{c}{28 VLBN models} \\ \midrule
		Method & \texttt{<0.5s} & \texttt{<2s} & \texttt{<10s} & \texttt{<1min} & \texttt{<10min} & \texttt{<1h} \\ \midrule
		\texttt{pyboolnet} & \texttt{0} & \texttt{0} & \texttt{0} & \texttt{0} & \texttt{0} & \texttt{0} \\
		\texttt{trapmvn} & \texttt{0} & \texttt{0} & \texttt{7} & \texttt{12} & \texttt{18} & \texttt{22} \\
		\fmpbn{} & \texttt{1} & \texttt{5} & \texttt{10} & \texttt{16} & \texttt{20} & \texttt{20} \\
		\lmpbn{} & \texttt{4} & \texttt{8} & \texttt{12} & \texttt{16} & \texttt{24} & \texttt{28} \\
		\midrule
		\multicolumn{7}{c}{100 AEON models} \\ \midrule
		Method & \texttt{<0.5s} & \texttt{<2s} & \texttt{<10s} & \texttt{<1min} & \texttt{<10min} & \texttt{<1h} \\ \midrule
		\texttt{pyboolnet} & \texttt{0} & \texttt{0} & \texttt{0} & \texttt{0} & \texttt{0} & \texttt{0} \\
		\texttt{trapmvn} & \texttt{0} & \texttt{14} & \texttt{56} & \texttt{96} & \texttt{100} & \texttt{100} \\
		\fmpbn{} & \texttt{99} & \texttt{100} & \texttt{100} & \texttt{100} & \texttt{100} & \texttt{100} \\
		\lmpbn{} & \texttt{100} & \texttt{100} & \texttt{100} & \texttt{100} & \texttt{100} & \texttt{100} \\
	\end{tabular}
\end{table}

\bibliographystyleSupp{unsrt} 
\bibliographySupp{ref}

\end{document}